\documentstyle[prl,twocolumn,aps,epsf]{revtex}
\begin{document}

\title{Anisotropy as a consequence of pre-equilibrium rescattering}

\author{Peter Filip}

\address{
Department of Theoretical Physics, Comenius University Bratislava \\ 
Institute of Physics, Slovak Academy of Sciences, Bratislava}

\maketitle

\begin{abstract} We show that azimuthal asymmetry in transverse momentum
distribution of particles observed in non-central heavy ion collisions
can originate from pre-equilibrium rescattering. 
The phenomenon is studied 
using computer simulation of expanding pion gas created in
Pb+Pb 160 GeV/n collisions.
Results obtained are discussed in comparison with experimental results of 
NA49 collaboration. Theoretical understanding of mechanism generating
the asymmetry is presented. Conclusion about non-equilibrium properties 
of the phenomenon studied is drawn. 
\end{abstract}


\section{Introduction}
\label{Into}

Since the discovery of directed flow in pioneering experiments on Bevalac
\cite{Bevalac} the phenomenon of azimuthal asymmetries has been studied
extensively at various beam energies using different types of colliding nuclei.
Besides the interesting phenomenon of directed flow two
additional signatures of the collective behavior of hadrons have been identified
in heavy ion collisions (HIC): 

1) {\it Radial flow}, 
existence of which has been clearly established in central collisions
\cite{RadialF}. Recently signature of radial flow in
{\it non-central} collisions has been predicted \cite{Voloshin} and most likely
also confirmed experimentally \cite{AMPQM97}. 

2) {\it Elliptic flow},
which can be oriented orthogonally to the reaction plane (squeeze-out) at
Bevalac energies \cite{Squeeze-out} or parallel to the reaction plane 
at AGS \cite{E877,Sorge} or SPS \cite{AMPQM97,APS97} energies. 
In Ni+Ni 2GeV/n experiments orientation of the elliptic flow depends on
transverse momentum region of neutral mesons selected for the analysis
\cite{Hlavac}.

Phenomenon of azimuthal asymmetries is satisfactorily understood as 
a consequence of collective hydrodynamical behavior of nuclear matter
described by the equation of state or as a result of
shadowing effect and absorption processes in spectator
parts of colliding nuclei.

In this work we study properties of second-order  elliptic flow of 
final state pions. We concentrate mainly on anisotropy  
created in Pb+Pb 160 GeV/n non-central collisions studied recently by NA49
collaboration using 
Main TPC \cite{AMPQM97}
and 
Ring Calorimeter \cite{TWQM96} 
data. 

Second-order flow oriented parallel to the reaction plane was predicted in
theoretical work of J.-Y.Ollitrault \cite{Ollie} for ultra-relativistic
energy domain. Mechanism of in-plane elliptic flow generation \cite{Ollie}
is based on the assumption of thermalization of nuclear matter in
the asymmetrical overlapping region of non-centrally colliding nuclei.
This assumption is essential for the existence of different pressure gradients
in different directions inside the asymmetrical overlapping region. 
During the subsequent hydrodynamical evolution 
different pressure gradients generate second-order
asymmetry in transverse momentum distribution of particles \cite{Ollie}. 
Also in the recent theoretical work on
elliptic flow at AGS energies \cite{Sorge} an assumption about
weak sensitivity of the resulting flow asymmetry to non-equilibrium
effects (in central rapidity region) is formulated.

In this work we show that elliptic flow can be generated by rescattering
process among produced hadrons (pions) 
without the equilibration in  prehadronic or  hadronic stage
of heavy ion collision. Our study is motivated by results of computer
simulation \cite{APS97} of the rescattering process
in the expanding pion gas created in Pb+Pb 160 GeV/n non-central collisions.

The paper is organized as follows: In section II we briefly describe our
rescattering simulation of the expanding pion gas - the CASCUS 1.0 model.
Main results of the simulation - centrality dependence, $p_t$ dependence,
rapidity dependence and the equilibration dependence of the 
asymmetry are presented in section III. The results are 
compared with experimental data \cite{AMPQM97}. 
Qualitative arguments offering our understanding of the mechanism 
generating the asymmetry are presented in Section IV.
At the end of this work 
a summary of results is given and short conclusions are drawn.

\section{Rescattering process in pion gas}
Total number of pions generated in ultrarelativistic Pb+Pb
collisions at SPS energy is surprisingly
large. Even more than 2500 pions can be generated in a central Pb+Pb collision
at this energy. Number of secondary particles thus exceeds number of
primary nucleons in the system. Assuming that pions are created in
independent nucleon-nucleon collisions the initial 
transverse momentum and rapidity spectra of produced pions are determined
mainly by the dynamics of pion production in nucleon-nucleon collisions
\footnote{Even if this picture of pion formation may be oversimplified 
we believe that it can provide a reasonable approximation to reality
for SPS energy region}.

Our rescattering simulation is based on the following scenario:
Initial state of the pion gas is characterized
by momentum and spatial distribution $\rho (\vec p,\vec x,t)$ of
pions as they are produced in nucleon-nucleon collisions.
As the system evolves positions of pions change naturally due to
the classical motion of pions. Momenta of pions can change as well in 
mutual collisions of pions.
As the size of the cloud of pions increases,
spatial density of pions becomes smaller and consequently collisions
among pions become rare. At some point collisions cease at all and
momentum distribution of pions is not influenced any more. Pions move then
freely on their way to detectors. Thus momentum distribution
of pions recorded by detectors is not equal to the original
momentum distribution of pions as they were produced in nucleon-nucleon
interactions. It is changed during rescattering process in the early
stage of the pion gas expansion.
In next subsection we give short description of the computer simulation
we have performed.

\subsection{Simulation of rescattering process}
Our rescattering model \cite{APS97} is similar to the
rescattering simulation described by T.J.Humanic \cite{Humanic}.
Two additional concepts have been used in our version of rescattering model:

1) Initial state of the pion gas in non-central collisions was generated
by cascade program \cite{Zavada} and 

2) Formation time of
hadrons \cite{FormationT} was introduced. The simulation is performed
in CMS of the Pb+Pb collision with time step $\Delta T=0.1fm/c$.
In each time step positions of pions are changed according to the
classical equation of motion
$\vec x(T+\Delta T)=\vec x(T)+\vec v\cdot \Delta T$. 
Two pions collide when their distance
is smaller than critical distance $d_c(s)=\sqrt{\sigma (s)/\pi}$ where
$\sigma (s)$ is isospin averaged elastic pion-pion cross section
(taken
from experimental data \cite{Estabrooks}).
New momenta of pions are determined in
CMS frame of pion-pion collision using differential cross section
of the elastic pion-pion interaction. 
Then new momenta of pions are transformed back
to the global frame of the simulation (CMS of Pb+Pb system).

\vskip0.5cm
\centerline{\epsfxsize=8cm\epsffile{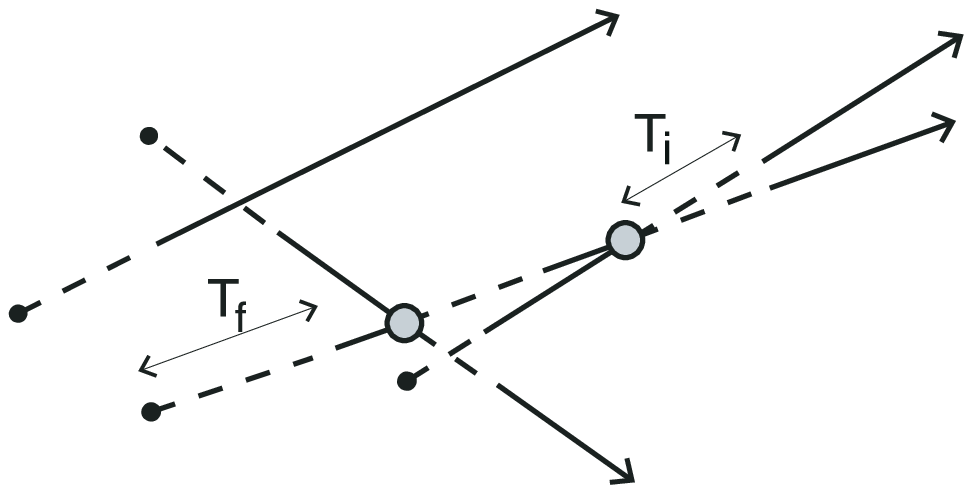}}
\vskip3.7pt
\centerline{\parbox{8cm} {\small {\bf Fig.1}
Time delay parameters incorporated in our rescattering simulation.
}}
\vskip0.4cm

After being created pions are not allowed to participate in mutual 
interactions for
Lorentz dilated time interval - the formation time
$T_f=\gamma \cdot \tau _f$ and for interaction delay time $T_i$ after
each collision. Formation time $\tau _f$ influences expansion of the
pion gas mainly in the initial stage of time evolution, interaction
time $T_i$ suppresses unphysical effects of rescattering simulations
\cite{PrattNonPhys} and  allows to influence total number of collisions
in the pion gas.

Since no collective behavior of primary nucleons is incorporated in
cascade generator \cite{Zavada} and since the production of pions in
primary nucleon-nucleon collisions is azimuthally isotropic the initial
state of the pion gas is azimuthally isotropic in momentum
space also in the case of non-central collisions:

\begin{equation}
\varrho(\vec x,\vec p,t)=\chi(t)\cdot A^T(\vec x) \cdot S^T(\vec p)
\label{psixpt}
\end{equation}
Here $A^T$ denotes azimuthally asymmetrical and $S^T$ denotes
azimuthally symmetrical distribution in transverse plane. During
the pion gas expansion symmetrical $S^T(\vec p)$ distribution
is changed by mutual collisions of pions.

First results of our simulation (centrality dependence of the asymmetry)
have been presented during Heavy Ion
Workshop on Particle Physics (September 1996) in Slovakia
\cite{APS97,CUPH1,LANL9605}. 
Detailed description of the
simulation can be found in thesis \cite{PhD}.
In following subsections we present additional
results of the rather extensive simulation we have performed.

\section{Results of the simulation}
Rescattering simulation was performed for random and unknown orientations 
of the reaction
plane in events generated by cascade program \cite{Zavada}. During the
subsequent analysis impact parameter orientation had been determined for
each event \cite{CUPH1} and histograms of azimuthal distributions of pions
had been filled (see Fig.2). Azimuthal asymmetry of second order oriented
in the reaction plane (not present in the initial state of the pion gas)
had been clearly identified in transverse momentum distribution of pions
after the rescattering process.

\vskip0.7cm
\centerline{\epsfxsize=8cm\epsffile{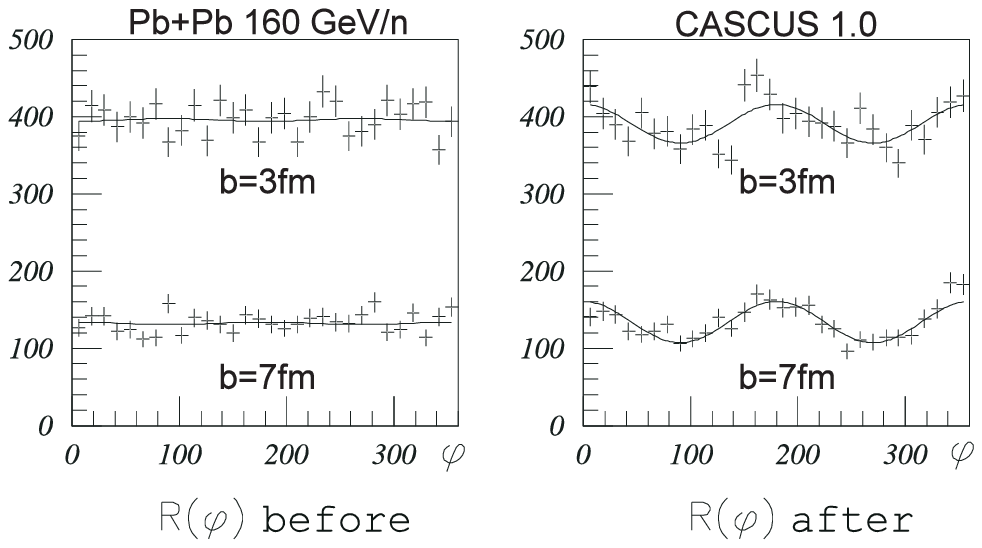}}
\vskip1.7pt
\centerline{\parbox{8cm} {\small {\bf Fig.2}
Azimuthal distribution of pions before and after the rescattering process for
events with $b=3$fm and $7$fm. 
}}
\vskip0.4cm
Strength of the asymmetry was evaluated
by fit to the function:
\begin{equation}
R(\phi )=S_o[1+S_2\cdot \cos (2\phi )]
\label{S2}
\end{equation}
In next subsections we present main properties of the asymmetry found.

\subsection{Centrality dependence}
Events with impact parameter $b\!=\!3,5,7,9,10,11,12$fm
generated by cascade program \cite{Zavada} have been used as input for the
rescattering program and subsequently analyzed for second-order
asymmetry. 
Following behavior had been found for 
different values of rescattering model parameters $\tau _f, T_i$:

\vskip0.7cm
\centerline{\epsfxsize=7.5cm\epsffile{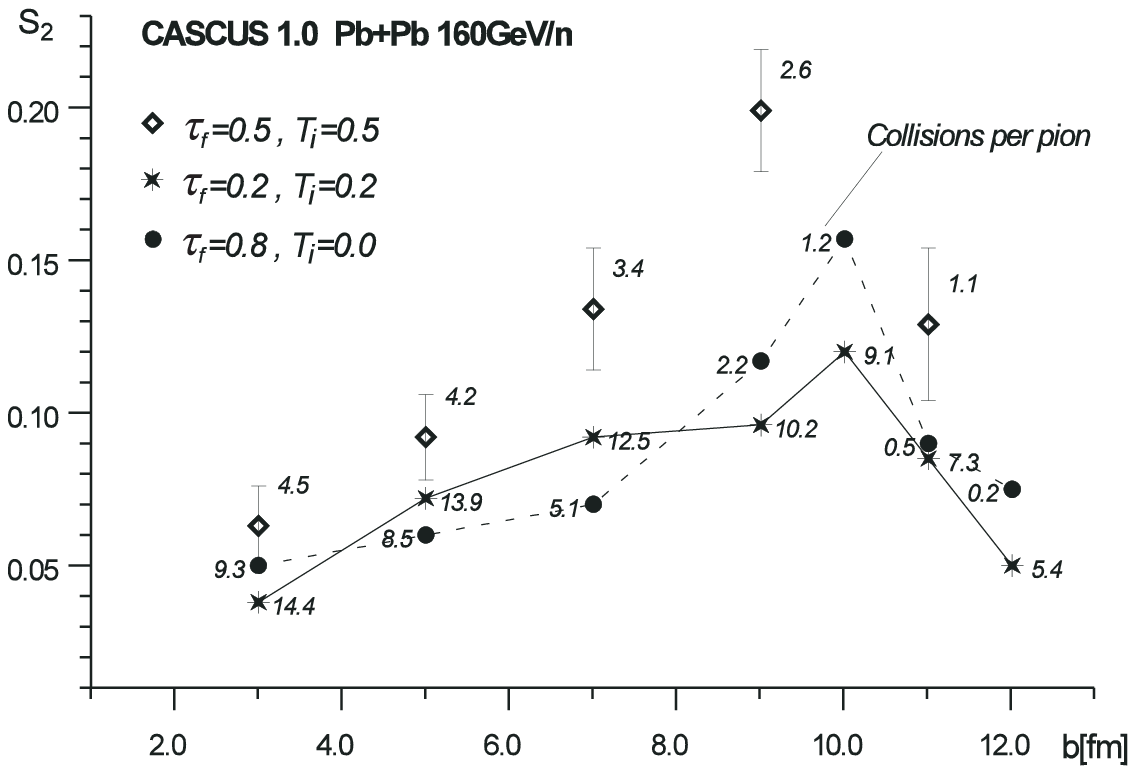}}
\vskip1.7pt
\centerline{\parbox{8cm} {\small {\bf Fig.3}
Centrality dependence of the asymmetry for different values of $\tau _f, T_i$
parameters. Number of collisions per pion (indicated for each data point)
decreases with increasing value of impact parameter. 
Non-zero asymmetry is
generated also for small number of collision per pion 
(0.5 Coll/pion at b=11fm). 
}}
\vskip0.4cm

Asymmetry increases with impact parameter up to the region $b=$9\,-10$fm$ and
then it decreases.
This behavior has been predicted in the work of J.-Y. Ollitrault \cite{Ollie}
and it is in remarkable agreement with experimental data \cite{TWQM96}.
From the point of view of rescattering simulation centrality dependence of
the asymmetry can be understood easily. For $b=0$ initial conditions
have cylindrical symmetry and there is therefore no reason for the
asymmetry in the final state. For very peripheral collisions interactions
of pions are very rare due to low multiplicity of pions. 
Therefore asymmetry is expected to decrease for peripheral collisions.
Sizeable asymmetry is generated in the intermediate region where spatial 
asymmetry in the initial state and number of
pion-pion collisions are sufficiently large.

Shape of the centrality dependence is not very sensitive to the number of
collisions per pion in the expanding pion gas (see Fig.3). 

We find also to be surprising, that the asymmetry is generated in the
case of small number of collisions per pion (e.g 0.5 collision per pion).
Dependence of the strength of asymmetry on the equilibration of the pion gas
is studied at the end of this section. Now we present results on
$p_t$ and rapidity dependence of the asymmetry.

\subsection{Transverse momentum dependence}
Transverse momentum dependence of the asymmetry has been identified in 
our results already in the early stage of this work \cite{APS97}.
It has been studied quantitatively for different values of
rescattering program parameters $\tau _f, T_i$. 
100 events with impact parameter $b=7fm$ have been used for this analysis.
Selected results are presented in Fig.4.

\vskip0.7cm
\centerline{\epsfxsize=7cm\epsffile{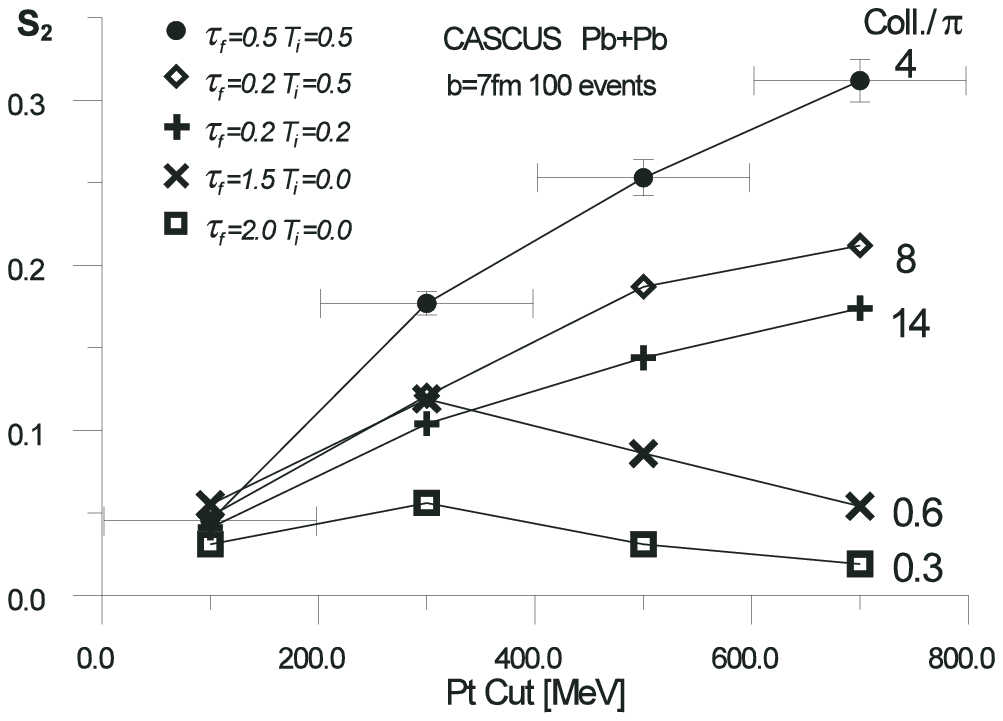}}
\vskip1.7pt
\centerline{\parbox{8cm} {\small {\bf Fig.4}
$p_t$ dependence of the elliptic flow of pions for different values
of parameters $\tau _f, T_i$.
}}
\vskip0.4cm

The asymmetry increases with $p_t$ for 
number of collisions per pion $N_{coll./\pi} = 4,8,14$ 
in rough agreement
with experimental data \cite{AMPQM97}.
For $N_{coll./\pi}=0.6,0.3$  the shape of the $p_t$ dependence
changes together with absolute strength of the asymmetry. Decrease
of the asymmetry strength at large $p_t$ for $N_{coll./\pi}=0.3,0.6$
can be understood easily. High $p_t$ pions can escape from the pion gas
volume without interaction due to large value of Lorentz-dilated 
formation time $T_f$
and also due to their larger velocity. Therefore smaller asymmetry is generated
for rarely colliding high-$p_t$ pions while larger asymmetry is 
generated for colliding
low-$p_t$ pions. Decrease of the asymmetry in case
$N_{coll./\pi}=8,14$ in comparison with $N_{coll./\pi}=4$ is discussed
in section {\bf D}.

Detailed comparison of $p_t$ dependence of asymmetry with experimental data
may allow to determine physical parameters incorporated in the simulation
(in particular the formation time).

For this goal however improvements in our rescattering model 
(e.g. dynamics of $T_i$ delay parameter) and also
higher statistics of experimental data \cite{AMPQM97} would be necessary.

\subsection{Rapidity dependence}
Rapidity dependence of the asymmetry has been studied using $b=7fm$
events sample. Resulting rapidity dependence found by our simulation
exhibits maximum at distance $\Delta Y = 0.75$ from central rapidity
region. 
We find to be interesting that results on rapidity dependence of elliptic
flow presented during Quark Matter 97 conference \cite{AMPQM97} also
exhibit decrease of the asymmetry strength at mid-rapidity region.

\vskip0.7cm
\centerline{\epsfxsize=7.3cm\epsffile{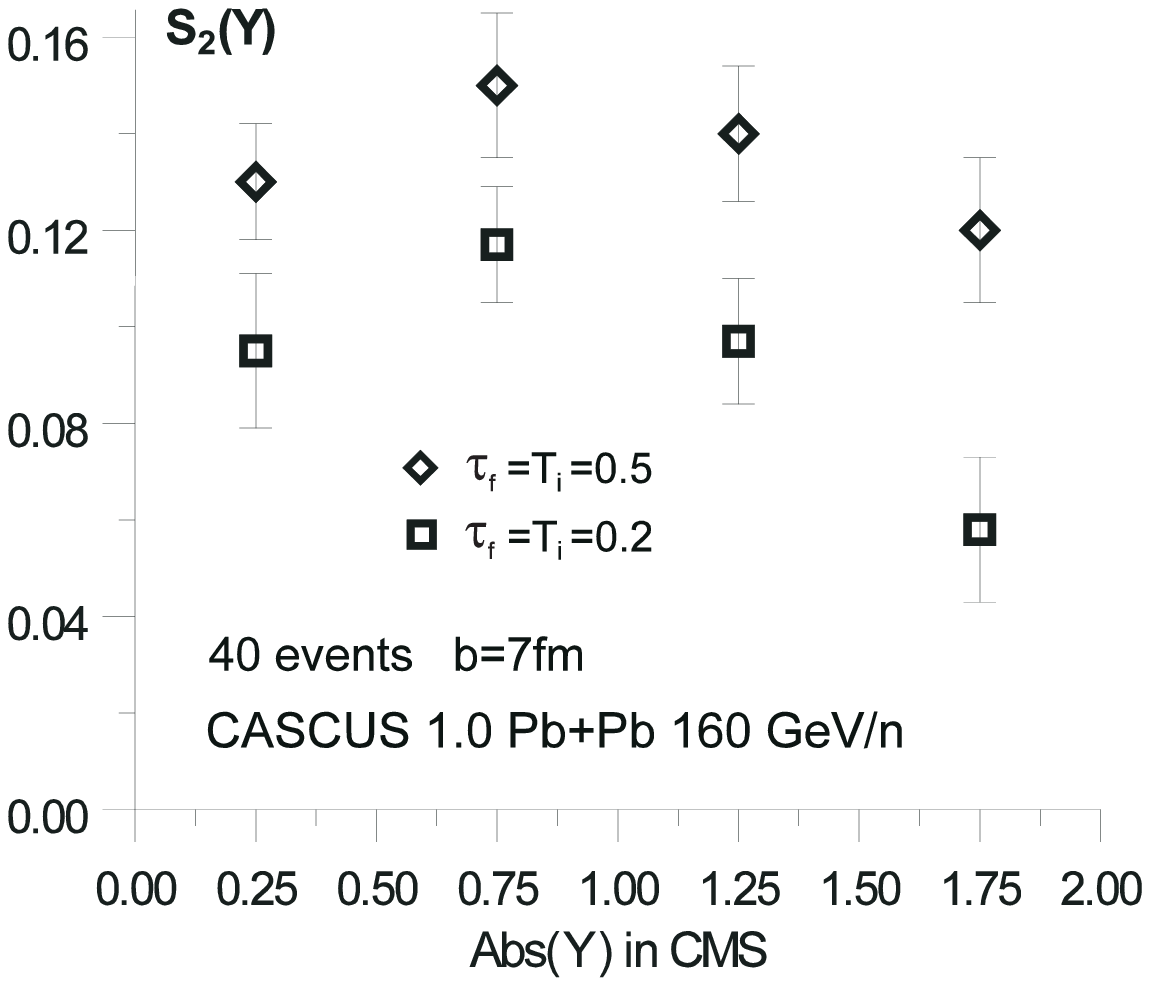}}
\vskip1.7pt
\centerline{\parbox{8cm} {\small {\bf Fig.5}
Rapidity dependence of the asymmetry. Note decrease of the
asymmetry at midrapidity.
}}
\vskip0.4cm

This result is worth of further investigation.
It is not excluded that the origin of the mid-rapidity decrease 
of the asymmetry
is related to behavior 
studied in the next subsection.

\subsection{Thermalization dependence}
Numerical value of the rescattering model parameters (see Fig.1)
allows to influence total number of collisions in the expanding pion gas.
We have studied dependence of the asymmetry on total
number of collisions in the pion gas for the set of $b=7fm$ events.
Following result has been obtained (Fig.6):

With increasing number of collisions in the pion gas the asymmetry first
increases, it exhibits maximum close to  $N_{coll/\pi}=4$ region
and then it decreases.  This decrease depends mainly on number of
collisions per pion, it is
not a consequence of the physics of formation time $\tau _f$ or interaction
time $T_i$ parameter. Both $\tau _f=0.5, T_i=0.2$ and 
$\tau _f=0.2, T_i=0.5 $ combinations lead to decrease of asymmetry from 
$\tau _f=0.5, T_i=0.5$ level (see Fig.6).
This result can be
interpreted as a non-equilibrium feature of the mechanism
generating the asymmetry. We think that monotonic increase and saturation
should be expected if the asymmetry were generated in an equilibrium process.

\vskip0.7cm
\centerline{\epsfxsize=7.8cm\epsffile{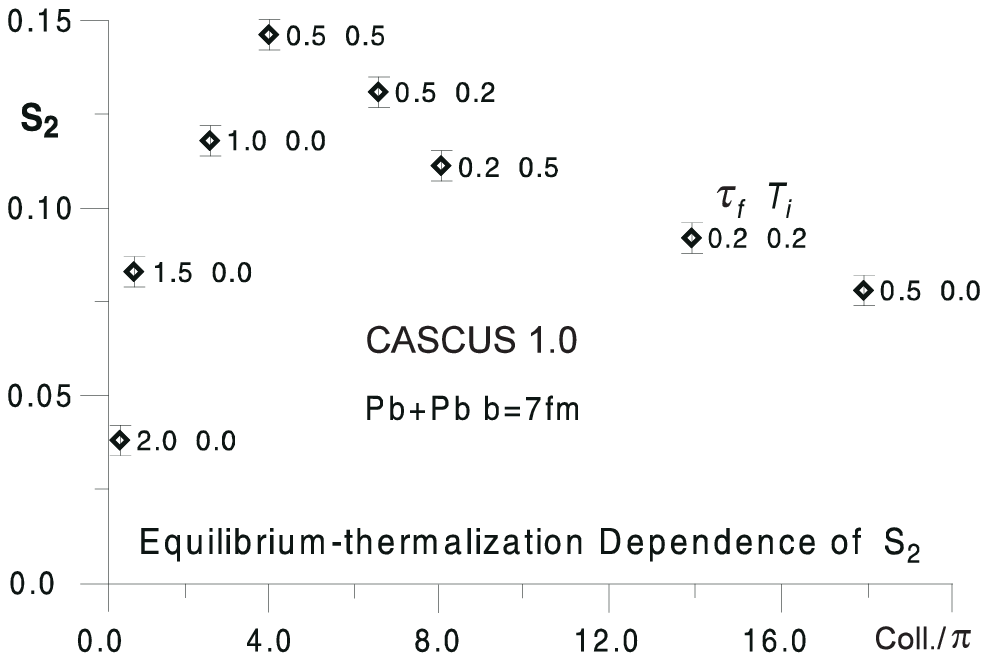}}
\vskip1.7pt
\centerline{\parbox{8cm} {\small {\bf Fig.6}
Thermalization dependence of the asymmetry.
}}
\vskip0.4cm

For high number of collision per pion 
$N_{coll./\pi} \approx 16$
the asymmetry seems to saturate.
We guess this corresponds to hydrodynamical limit 
studied in work \cite{Ollie}.  

At this point we can try to understand the 
decrease of asymmetry at mid-rapidity found
in our simulation (Fig.5) and present also in experimental data \cite{AMPQM97}.
The decrease can be explained as a consequence of smaller number of 
collisions per pion ($N_{coll./\pi}<4$) at
midrapidity  
in comparison with 
number of collisions per pion at the maximum of rapidity 
dependence\footnote{This explanation is supported by  
the simulation of
dilepton production from pion gas \cite{Jupi}.
}. 
This point requires further and more careful study since
the position of the maximum is different in our results and 
experimental data \cite{AMPQM97}.

\subsection{Time evolution of the asymmetry}
Numerical simulation of the pion gas expansion allows us to study also
quantities which cannot be accessed from
final stage of the pion gas.
We have studied time evolution of the number of collisions
in the pion gas together with the time evolution of second-order asymmetry.
For the sake of simplicity the asymmetry
has been characterized by $R_p$ parameter: 

\begin{equation}
R_p=\frac{\langle p_y^2\rangle }{\langle p_x^2\rangle }
\label{Rp}
\end{equation}
which does not require the fit procedure for its evaluation.

From results shown in Fig.7 we conclude that the asymmetry is generated
during the early stage of the expansion. This confirms our
conjecture that the asymmetry can originate during pre-equilibrium
stage of the pion gas expansion.
Theoretical understanding of mechanism the asymmetry is generated by is
presented in next section.

\vskip0.7cm
\centerline{\epsfxsize=8cm\epsffile{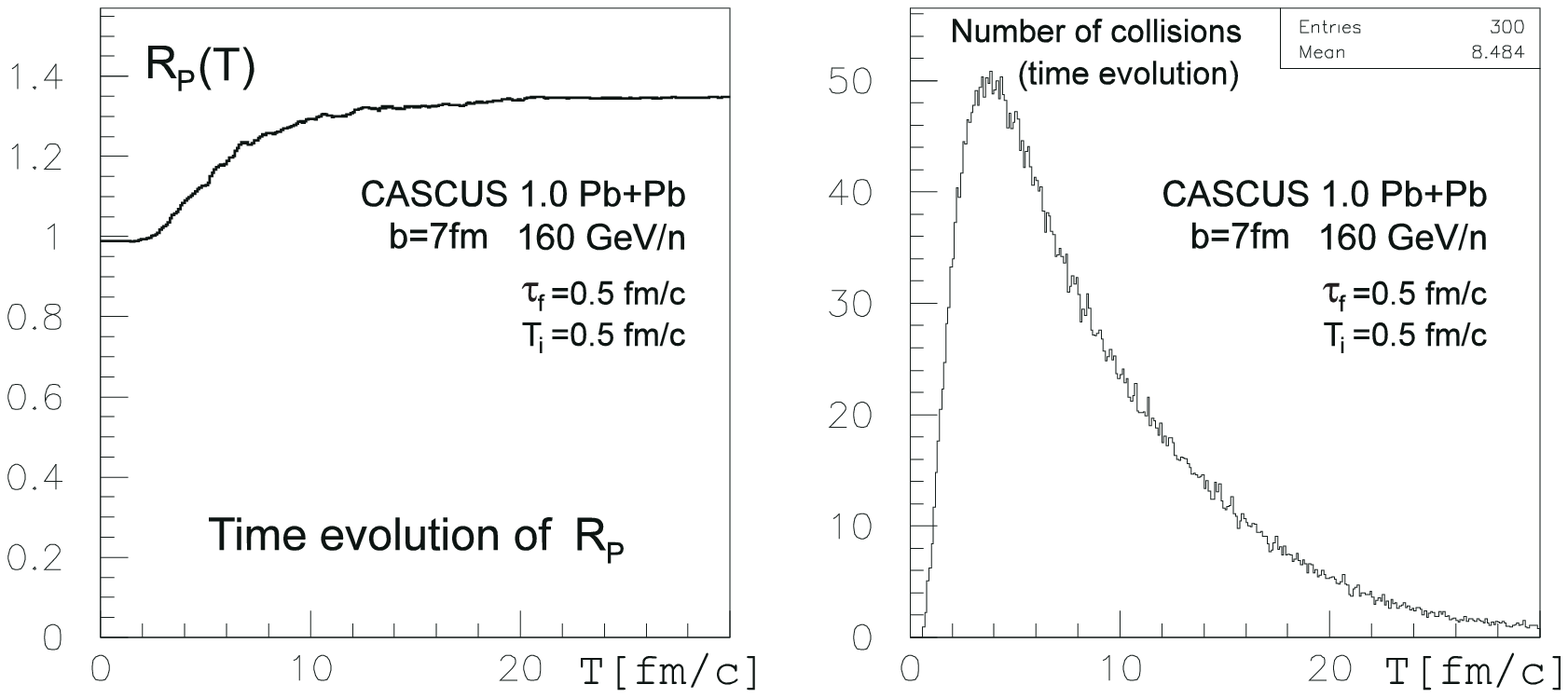}}
\vskip1.7pt
\centerline{\parbox{8cm} {\small {\bf Fig.7}
Time evolution of the symmetry parameter $R_p(t)$ and evolution of
number of collisions during the expansion of pion 
gas.
Time dependence of $R_p(T)$  is nearly identical with
the behavior of $S_2(T)$ parameter \cite{PhD}.
}}
\vskip0.4cm

\section{Theoretical analysis of asymmetry}
Expansion of the interacting pion gas can be described by
kinetic equation \cite{Lifshics X} which (in the case of zero external
macroscopic field) takes the form:

\begin{equation}
\frac {\partial f(\vec p, \vec x,t)}{\partial t}+
       \vec v_{p}\cdot \nabla _{\vec r} f(\vec p, \vec x,t) =
       \left[{St}\right]_{Coll}
\label{Kinet}
\end{equation}
$\left[ St\right]_{Coll}$ is the collision (or scattering)
term describing influence of pion-pion collisions on distribution
function $f(\vec p,\vec x,t)$.

In slightly simplified case
(we do not take into account time interval during which pions are produced)
our initial state of the pion gas - Eq.(\ref{psixpt}) can be written in
 the form:

\begin{equation}
\rho (\vec x,\vec p,T_0)=A^T(\vec x_t,x_z)\cdot S^T(\vec p_t,p_z)
\label{InitX}
\end{equation}
where again $A^T(\vec x_t,x_z)$ denotes asymmetrical spatial distribution
of pions in transverse plane and $S^T(\vec p_t,p_z)$ is symmetrical
distribution of pions in transverse momentum.
During the process of pion gas expansion asymmetry in spatial distribution
of pions ''leaks'' into momentum distribution of pions
which becomes asymmetrical in transverse momentum space.

This phenomenon
results from mutual interactions of pions in the expanding asymmetrical
volume. If there were no collisions of pions, momentum distribution
of pion gas (momenta of pions) would be unchanged.
This is clear intuitively and it can be seen also on analytical level
after the integration of kinetic equation (\ref{Kinet}) in spatial coordinates
and after the subsequent time integration:

\begin{equation}
F_A(\vec p_t,p_z,T)=
F_S(\vec p_t,p_z,T_0) + \int _{T_0}^T\!\!\int \left[ St \right] d^3xdt
\label{KinT}
\end{equation}

Here distribution $F(\vec p,t)=\int F(\vec p,\vec x,t)d^3x$ is distribution
of pions in momentum space.
Since original momentum distribution of pions $F_S(\vec p,T_0)$ is
symmetrical in $\vec p_t$ plane (our initial condition) and since resulting
distribution $F_A(\vec p,T)$ is asymmetrical (this distribution is measured
by detectors) the collision integral
$\int _{T_0}^T\int \left[St\right]d^3\!\! xdt $
must be asymmetrical.

Asymmetry of collision integral in Eq.(\ref{KinT}) originates from
the asymmetrical spatial distribution of pions in transverse
plane. We do not attempt to evaluate the collision integral here though
we guess this is accomplishable task. The asymmetry of the collision
integral can be  understood intuitively using Toy model of pion gas expansion
described in the previous paper \cite{APS97}.
In next section
we derive formalism of two point decomposition of source which
(we think) allows to study the asymmetry theoretically.

\subsection{Two point source of pions.}
Let us consider two point-like sources of pions separated by distance $D$
in transverse plane. Each source radiates pions with random azimuthal
angle and (for simplicity) with fixed absolute value of momentum.
If we switch off rescattering of pions emitted from these
two sources azimuthal distribution of pions observed in a distant detector
surrounding the source ($D\ll R$) is constant - symmetrical.
However if collisions of pions are allowed the resulting distribution
of pions is asymmetrical in transverse momentum plane. Excess of 
pions in direction orthogonal to $\vec D$ 
is generated as a consequence of collisions (see Fig.8).

\vskip0.2cm
\centerline{\epsfxsize=6.7cm\epsffile{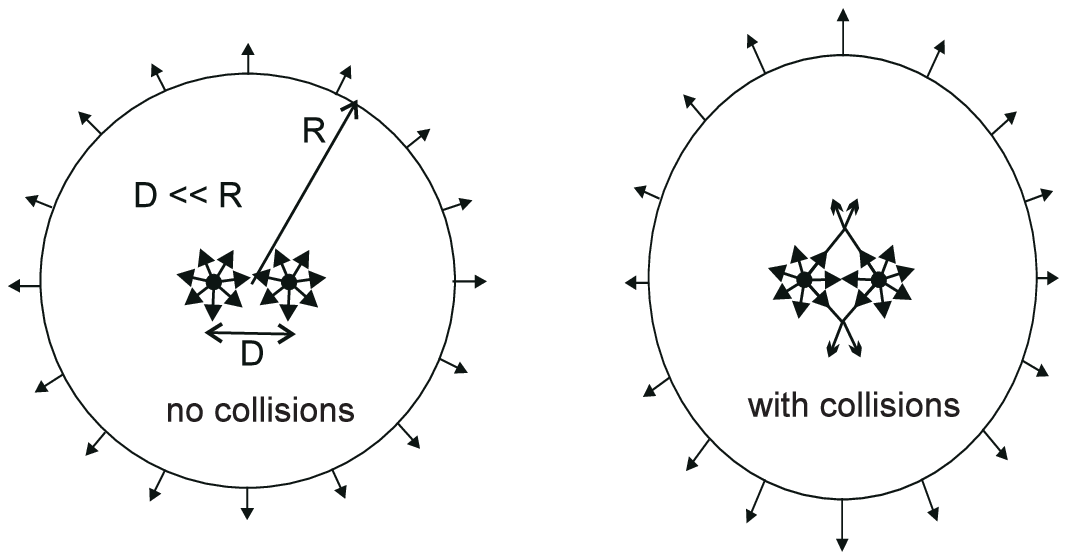}}
\vskip1.7pt
\centerline{\parbox{8cm} {\small {\bf Fig.8}
Azimuthal asymmetry generated by interactions of pions emitted from two
point-like sources of pions.
}}
\vskip0.4cm

Elliptic asymmetry can be characterized by $R_p$
parameter - Eq.(\ref{Rp}) and therefore we shall express
$p_{y'_1}^2+p_{y'_2}^2$ quantity for a given pair of
pions after the collision at any place of transverse plane.
If $p_{y'_1}^2+p_{y'_2}^2$ has increased 
then also $R_p$ parameter increases what means that elliptic asymmetry
is generated. 
After Lorentz boosting back from CMS of $\pi\pi $  collision 
to the rest frame of two-point source and after averaging over random
orientations of pion momenta after the collision (in CMS of $\pi\pi$ collision) 
we obtain the result (see Appendix):

\begin{equation}
\langle p_{y'_1}^2+p_{y'_2}^2\rangle _{\alpha} 
 = p_{C\!M\!S\!}^2 + \gamma ^2\beta _y^2
    \cdot \Lambda (p_{C\!M\!S\!},\beta)
\label{PxyA}
\end{equation}
where 
$\Lambda (p_{C\!M\!S\!},\beta)=2E_{C\!M\!S\!}^2+p_{C\!M\!S\!}^2\left [
      \frac{2}{\gamma +1}+\frac{\gamma ^2\beta ^2}{(\gamma + 1)^2} \right ]
$ and 
$\vec \beta = \frac {\vec p_1 + \vec p_2}{E_1+E_2}$. For 
$\langle p_{x'_1}^2+p_{x'_2}^2\rangle _{\alpha}$ the same expression can be 
found just $\beta _y$
is replaced by $\beta _x$.
Increase of $R_p$ parameter is generated by positive values of
\begin{equation}
\Delta p_y^2=\langle p_{y'_1}^2+p_{y'_2}^2 \rangle _{\alpha} 
- p_{y_1}^2-p_{y_2}^2
\label{delta}
\end{equation}
quantity ($p_{y_1},p_{y_2}$ are momenta of pions before collision) and negative
values of similarly defined $\Delta p_x^2$ quantity. Numerical
values of $\Delta p_y^2 $ and $\Delta p_x^2$ depend
 on coordinates of the collision point
in transverse plane since $\vec \beta, p_{C\!M\!S}$ 
and also $\vec p_1, \vec p_2 $
depend on coordinates of the collision point.
In Fig.9 we show Mathematica \cite{Math} contour-plot of $\Delta p_y^2$
quantity for two point source radiating pions with equal ($p_1=p_2$) and
different ($p_1>p_2$) momenta. 

Two-point source represents (in rough approximation) overlapping region
of colliding nuclei in transverse plane. 
It is clear, that resulting elliptic asymmetry asymmetry is
oriented in the direction of impact
parameter - orthogonally to $\vec D$ vector.

\vskip0.7cm
\centerline{\epsfxsize=8cm\epsffile{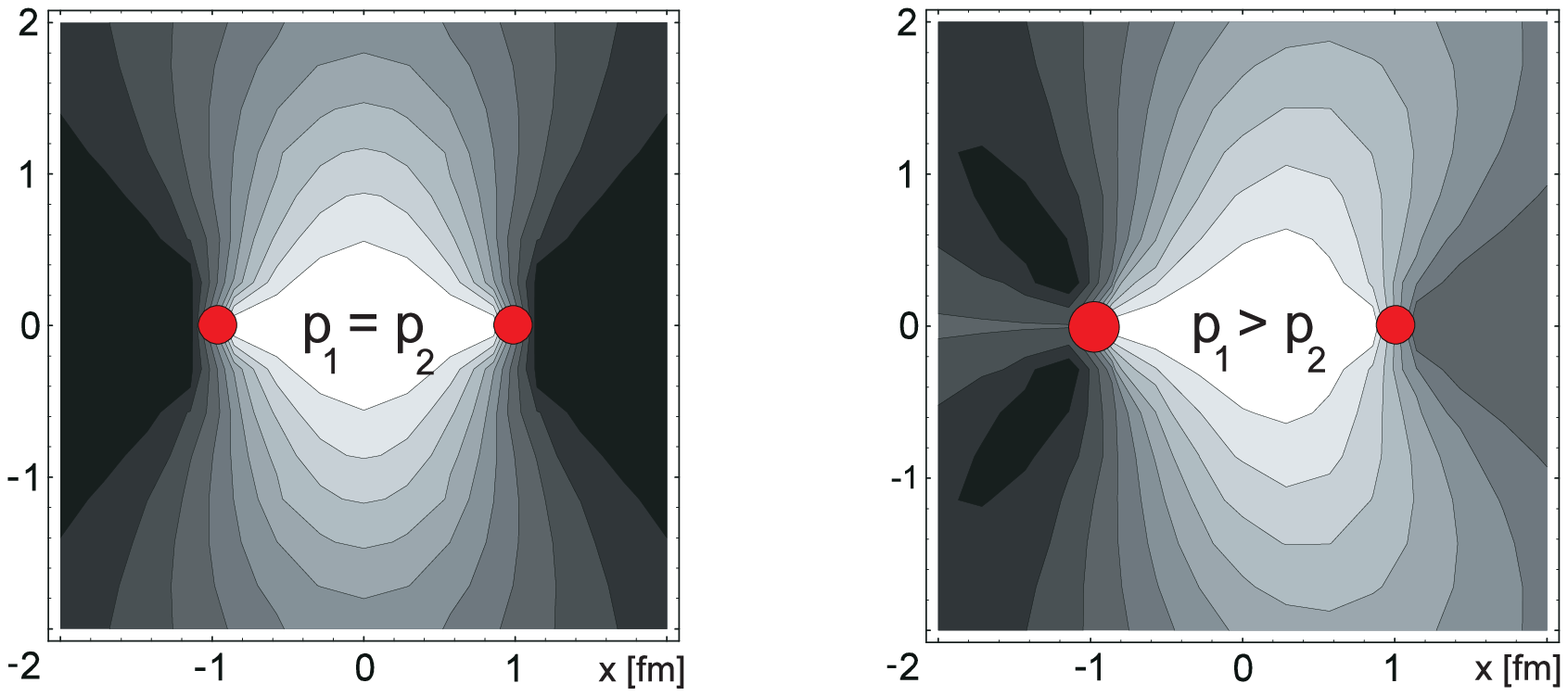}}
\vskip5.7pt
\centerline{\parbox{8cm} {\small {\bf Fig.9}
Contour plot of 
$
\Delta p_y^2$ - see Eq.(\ref{delta})
for a two point source emitting pions with momenta $p_1$ and $p_2$. 
White color represents positive values of $\Delta p_y^2$ quantity.
}}
\vskip0.4cm

In next section we study
asymmetry generated by non-trivial discrete and also continuous source.

\subsection{Two-point source decomposition}

Main idea of the
two-point decomposition of source is following: Each spatially
distributed source of interacting particles
behaves 
(in 1 collision/pion approximation) as a set of two-point sources
(see Fig.10).
Each two-point source (characterized by vector $\vec D$) 
generates asymmetry with the orientation determined by vector 
$\vec D$. Our intention is to express analytically  resulting 
asymmetry generated by the whole source.

Let $R_2(\phi ,\vec D)$ is azimuthal distribution of pions
(particles) emitted from two-point source characterized by vector
$\vec D$. Asymmetry of $R_2(\phi ,\vec D)$ is generated
by collisions of pions at any place of transverse plane. Contribution from
all possible collision points is averaged. 
Let us also have non-trivial
e.g. discrete source radiating independent pairs of interacting 
pions. (This is our specification of 1 collision per pion approximation.)

\vskip0.7cm
\centerline{\epsfxsize=8cm\epsffile{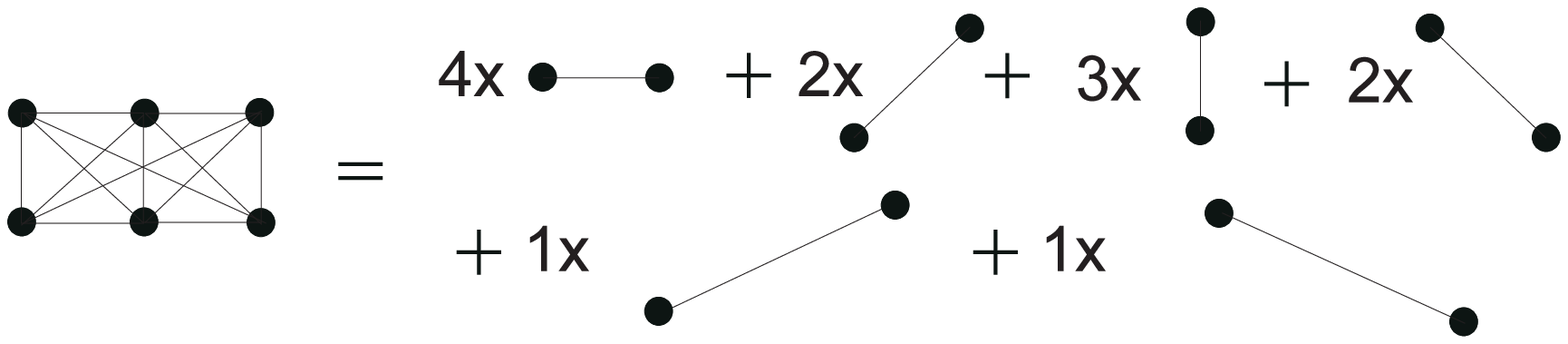}}
\vskip1.7pt
\centerline{\parbox{8cm} {\small {\bf Fig.10}
Discrete asymmetrical source of pions and its two-point decomposition.
Note the excess of horizontally oriented two-point sources in comparison
with vertical two-point sources.
}}
\vskip0.4cm

Then in
1coll./$\pi $ approximation 
resulting azimuthal asymmetry of particles emitted from more complex but
discrete source (see Fig.10) is given by weighted sum of azimuthal
distributions of all possible two-point sources:

\begin{equation}
R_{\rho }(\phi )=\sum _{ij} W_{ij}\cdot R_2(\phi ,\vec D_{ij})
\label{Summ1}
\end{equation}

Weights $W_{ij}$ depend on intensity of point-like sources $i,j$.
Expression (\ref{Summ1}) can be extended to the case of spatially
continuous source. 

If we define 
probability distribution for the emission of a pair of particles
from any places $\vec x_1,\vec x_2$ in the source $\rho (\vec x)$ 
at relative distance
$\vec D=\vec x_1-\vec x_2$ 

\begin{equation}
W_{\rho} (\vec D)=
          \int \!\!\! \int d^2x_1d^2x_2\left[\rho (\vec x_1)\rho (\vec x_2)
	  \cdot \delta (\vec D-(\vec x_1-\vec x_2))\right] 
\end{equation}
what can be rewritten 
as
\begin{equation}
W_{\rho}(\vec D) =\int \rho (\frac {\vec x+\vec D}{2})\cdot
		\rho (\frac {\vec x-\vec D}{2}) d^2\vec x
\label{SD}
\end{equation}
then resulting azimuthal distribution of particles emitted by
spatially distributed source $\rho (\vec x)$ can be expressed
in 1 collision per pion approximation as follows:

\begin{equation}
R_{\rho }(\phi )=\int W_{\rho}(\vec D)\cdot R_2(\phi,\vec D) d^2\vec D
\label{ThatIsIt}
\end{equation}

Dependence of the resulting $R_{\rho }(\phi )$ distribution on e.g.
formation time and $\pi \pi$ cross section is encoded in distribution
$R_2(\phi ,\vec D)$.
If properties of $R_2(\phi ,\vec D)$ distribution are known,
expression (\ref{ThatIsIt}) allows us to say which source can generate
asymmetry at analytical level - without numerical simulation.

Asymmetry of resulting $R_{\rho}$ distribution is generated only by
asymmetrical part of $W_{\rho }(\vec D)$ distribution
\begin{equation}
A(D,\phi )=W_{\rho}(D,\phi )-S(D) 
\end{equation}
where $S(D)$ is symmetrical part of $W_{\rho}(D,\phi)$. 
$S(D)$ is equal to minimum of $W_{\rho }(D,\phi )$
at given $D$. Asymmetrical source distribution $\rho (\vec x)$ leads
to asymmetrical $W_{\rho }(\vec D)$ distribution which results in
non-zero asymmetrical distribution $A(D,\phi)$. Non-zero $A(D,\phi)$
distribution generates asymmetrical $R_{\rho }(\phi )$. Properties
of $R_{\rho }(\phi )$ depend on analytical properties of $A(D,\phi )$ 
and $R_2(\phi, \vec D)$.

\section{Summary and conclusions}
We have shown that second order azimuthal asymmetry of hadrons known also as
in-plane elliptic flow can be generated by rescattering process.
We have performed extensive study of this asymmetry
using computer simulation of Pb+Pb 160 GeV/n non-central collisions.
Main properties of the asymmetry (impact parameter dependence, $p_t$
dependence and rapidity dependence) studied by the simulation are in
rough agreement with experimental data \cite{AMPQM97}.  

Transverse momentum dependence
of the asymmetry (Fig.4) is found to be sensitive to formation time parameter.
Careful analysis of experimental data together with improved  simulation
of hadron gas expansion might allow to determine experimental 
value of the formation time of pions from data.

Rapidity dependence of the second-order asymmetry of pions (Fig.5)
exhibits minimum at midrapidity.
Study of thermalization dependence
of the asymmetry (Fig.6) allowed us to interpret this behavior as a 
consequence
of different number of collisions among pions in different rapidity intervals.

Dependence of the strength of asymmetry on total
number of collisions in the pion gas (Fig.6) together with non-zero value
of resulting asymmetry in the $N_{coll/\pi}$=0.5 case (Fig.3) 
instigates us to conclude
that the studied asymmetry can be generated in a non-equilibrium
(pre-equilibrium) process. This conclusion substantially modifies
our point of view on physical meaning of the in-plane elliptic
flow of pions measured in experimental data.

We have developed formalism of two-point
source decomposition (in one collision per pion approximation) which
allows theoretical study of the asymmetry. We have shown that
resulting distribution of pions in momentum space is related to
the initial spatial distribution of pions. 

We conclude that azimuthal anisotropy in momentum distribution of
hadrons can be generated by pre-equilibrium
rescattering process. 

\section{Appendix}

For a given pair of pions colliding at place $(x,y)$ of transverse plane
the absolute value of momenta in CMS frame of collision is:

\begin{equation}
p_{C\!M\!S\!}=\sqrt{M(x,y)^2/4-m_{\pi}^2}
\end{equation}
where $M(x,y)^2=(E_1+E_2)^2-(\vec p_1(x,y)+\vec p_2(x,y))^2$.
Let the orientation of momentum
$\vec p^{\ '}_{C\!M\!S\!}$ after the collision in CMS of $\pi\pi$ collision
is random and characterized by angle 
$\alpha$: $p^{\ 'C\!M\!S\!}_{x_1}=p_{C\!M\!S\!}\cdot \cos (\alpha);
p^{\ 'C\!M\!S\!}_{y_1}=p_{C\!M\!S\!}\cdot \sin (\alpha)$. Then momentum
of pion after the collision in the LAB frame (rest frame of two-point source)
is:

\begin{equation}
\vec p_1^{\ '}=\vec p_1^{\ 'C\!M\!S\!}+
                \gamma \vec \beta '(\frac{\gamma}{\gamma +1}
\vec \beta '\cdot \vec p_1^{\ 'C\!M\!S\!} - E_{C\!M\!S\!})
\end{equation}
where $\vec \beta ' =-\vec \beta = - \frac{\vec p_1+\vec p_2}{E_1+E_2}$.
Consequently for e.g. $y$ components of pion momenta after collision we have:

\begin{eqnarray}
p_{y_1}^{\ '}=p^{\ 'C\!M\!S\!}_{y_1}+\gamma \beta _y \cdot G_1 \nonumber \\
p_{y_2}^{\ '}=p^{\ 'C\!M\!S\!}_{y_2}+\gamma \beta _y \cdot G_2
\label{Labxy}
\end{eqnarray} 
where $G_1=\frac{\gamma }{\gamma +1}\vec \beta 
           \cdot \vec p^{\ 'C\!M\!S\!}_1+E_{C\!M\!S\!}$
and
similarly 
$G_2=
\frac{\gamma }{\gamma + 1}\vec \beta 
 \cdot \vec p^{\ 'C\!M\!S\!}_2+E_{C\!M\!S\!}$.
Then the sum $p_{y_1}^{'2}+p_{y_2}^{'2}$ gives:

\vskip0.2cm
\begin{equation}
p_{y_1}^{'2}+p_{y_2}^{'2}=
2p^{'C\!M\!S\!\ 2}_{y_1}+\gamma ^2\beta _y^2G^{+2}_{12}
 +2p^{'C\!M\!S\!}_{y_1}\ \gamma \beta _y G^-_{12}
\label{fixeda}
\end{equation}

\vskip0.13cm
where $G^-_{12}=G_1-G_2$; $G^{+2}_{12}=G_1^2+G_2^2$  and relationship 
$\vec p^{\ 'C\!M\!S\!}_1=-\vec p^{\ 'C\!M\!S\!}_2$
has been
used. Equation (\ref{fixeda}) is valid for any given angle $\alpha $ of pion
momentum after the collision in CMS. For the sake of simplicity we assume
that this angle is random what allows to perform averaging of result 
(\ref{fixeda})
over values of $\alpha $. This leads to the result:

\begin{eqnarray}
\langle p_{y_1}^{\ '2}+p_{y_2}^{\ '2} \rangle _{\alpha }&=&
p^2_{C\!M\!S\!} + \gamma ^2 \beta _y^2
\left[ (\frac{\gamma}{\gamma +1})^2\beta ^2 p_{C\!M\!S\!}^2 
+2E_{C\!M\!S\!}^2\right ]
\nonumber \\
&+& \frac{2\gamma ^2}{\gamma +1} p_{C\!M\!S}^2 \beta _y^2
\label{FX}
\end{eqnarray}
where following sub-results have been used:
\begin{eqnarray}
\langle G_1^2+G_2^2 \rangle _\alpha=
2E_{C\!M\!S\!}^2 + \frac{\gamma ^2\beta ^2}{(\gamma +1)^2} p_{C\!M\!S\!}^2 \\
\langle \sin \alpha \cdot [G_1-G_2] \rangle _\alpha=
2p_{C\!M\!S\!} \frac {\gamma}{\gamma +1}\beta_y
\end{eqnarray}
Result (\ref{FX}) can be rewritten directly in the form of
Eq.(\ref{PxyA}).

\section{Acknowledgments}
The author is indebted to prof. J.Pi\v s\'ut for careful reading of this
manuscript and important comments and suggestions. 
This work was supported by Comenius University Bratislava, Open Society Fund
in Bratislava and Institute of Physics, Slovak Academy of Sciences under 
Grant No. 2/4111/97.

I am grateful to RIKEN-BNL center for the hospitality and financial support
during symposium on Non-equilibrium Many Body Physics organized in
BNL, September 1997.

\end{document}